\documentclass[conference]{IEEEtran}
\IEEEoverridecommandlockouts
\usepackage{cite}
\usepackage{amsmath,amssymb,amsfonts}
\usepackage{algorithmic}
\usepackage{graphicx}
\usepackage{textcomp}
\usepackage{xcolor}
\usepackage{todonotes}
\usepackage{xspace}
\usepackage[absolute,overlay]{textpos}

\def\BibTeX{{\rm B\kern-.05em{\sc i\kern-.025em b}\kern-.08em
    T\kern-.1667em\lower.7ex\hbox{E}\kern-.125emX}}
\begin{document}

\begin{textblock*}{20cm}(1cm,1cm) 
{\color{red}{This is the authors' version of the work. The final version of this paper will appear at IEEE ITSC 2022.}}
\end{textblock*}

\title{Securing Automotive Architectures with Named Data Networking}

\author{\IEEEauthorblockN{Zachariah Threet}
\IEEEauthorblockA{\textit{TNTech} \\
}
\and
\IEEEauthorblockN{Christos Papadopoulos}
\IEEEauthorblockA{\textit{The University Of Memphis} \\
}
\and
\IEEEauthorblockN{William Lambert}
\IEEEauthorblockA{\textit{TNTech} \\
}
\and
\IEEEauthorblockN{Proyash Podder}
\IEEEauthorblockA{\textit{Florida International Institute} \\
}
\and
\IEEEauthorblockN{Spiros Thanasoulas }
\IEEEauthorblockA{\textit{The University of Memphis} \\
}
\and
\IEEEauthorblockN{Alex Afanasyev}
\IEEEauthorblockA{\textit{Florida International Institute} \\
}
\and
\IEEEauthorblockN{Sheikh Ghafoor}
\IEEEauthorblockA{\textit{TNTech} \\}
\and
\IEEEauthorblockN{Susmit Shannigrahi}
\IEEEauthorblockA{\textit{TNTech} \\
}
}
\maketitle

\begin{abstract}

As in-vehicle communication becomes more complex, the automotive community is exploring various architectural options such as centralized and zonal architectures for their numerous benefits. Zonal architecture reduces the wiring cost by physically locating related operations and ECUs near their intended functions and the number of physical ECUs through function consolidation. Centralized architectures consolidate the number of ECUs into few, powerful compute units. Common characteristics of these architectures include the need for high-bandwidth communication and security, which have been elusive with standard automotive architectures. Further, as automotive communication technologies evolve, it is also likely that multiple link-layer technologies such as CAN and Automotive Ethernet will co-exist. These alternative architectures promise to integrate these diverse sets of technologies. However, architectures that allow such co-existence have not been adequately explored.

In this work we explore a new network 
architecture called Named Data Networking (NDN) to achieve multiple goals: 
provide a foundational security infrastructure and  bridge different link layer protocols such as CAN, LIN, and automotive Ethernet into a unified communication system.

We created a proof-of-concept bench-top testbed using CAN HATS and Raspberry PIs that replay real traffic over CAN and Ethernet to demonstrate how NDN can provide a secure, high-speed bridge between different automotive link layers. We also show how NDN can support communication between centralized or zonal high-power compute components. Security is achieved through digitally signing all Data packets between these components, preventing unauthorized 
ECUs from injecting arbitrary data into the network. We also demonstrate NDN's ability to prevent DoS and replay 
attacks between different network segments connected through NDN.


\end{abstract}


\section{Introduction}

As the requirements for in-vehicle communication increase in terms of higher bandwidth, isolation, and security requirement all while attempting to control costs, the automotive community is looking at various communication architectures including centralized and zonal architectures. 
The former centralizes functions but has a higher cost due to increased wiring requirements. The latter reduces the wiring requirements by moving physically related functions into zones, while using high-speed interconnects between zones such as automotive Ethernet.  
Zonal architectures connected via high-speed networks also enable isolation and new security mechanisms to protect zones. A zonal architecture simplifies in-vehicle network security by separating different functionality in separate buses\cite{ullah2021} and providing a convenient location to manage and filter traffic between buses.  In the short term, in-vehicle networks will continue to employ standard layer-2 technologies such as CAN (controller area networks) and CAN-FD (CAN-flexible data), but also are expected to use high-speed networks such as automotive Ethernet to interconnect zones.

Regardless of the model, we anticipate several commonalities in future automotive environments. These include the requirement for built-in security so that connected cars or individual components are not easily compromised. Future vehicles will also require high-speed, high-bandwidth communication networks primarily driven by entertainment and real-time sensors and cameras.   Finally, as the industry moves to newer hardware and features, it is also likely that multiple link-layer technologies such as CAN and Automotive Ethernet will co-exist.

As the automotive community explores new networking architectures for this transition, it is tempting to adopt well-tested and proven technologies such as the TCP/IP architecture.  However, TCP/IP has several well-known security limitations and this work argues that the automotive industry should investigate other networking architectures besides IP as they move away from existing architectures such as CAN\cite{9644625}. Specifically, we utilize \emph{Named Data Networking (NDN)}, an 
architecture that incorporates unified security-by-design from the network to the application layers.


NDN is an 
architecture that incorporates unified security-by-design from the network to the application layers. While NDN has not yet been used for in-vehicle communication, we argue is that it is superior to IP, especially in security, making it a strong candidate for in-vehicle communication. Unlike IP, which secures the communication channel between two entities, NDN secures the content through digital signatures that cryptographically bind a name to the content, ensuring both authentication and integrity of the data. 
This will also lead to performance gains because data packets can be precreated before a request is made for them. Further, packets are cached, this means that subsequent requests can be made for the same data packet without it having to be recreated. NDN also supports native multicast and can support efficient pub-sub models. Further, it can also be implemented directly over L2 or L3 layers.

In this work, we utilize a bench-top testbed to demonstrate the multiple advantages of NDN. Specifically, we demonstrate NDN is able to 
secure data by providing security by design where all Data packets are signed by the publisher that allows the data consumers to validate the packets before accepting them, and provide a unified communication system between CAN and Ethernet. We note that while our gateways only interface between two link-layer technologies (CAN and NDN), there is no architectural limitation on adding other technologies such as automotive Ethernet to the gateway.

We utilize the testbed not only to demonstrate connectivity but also security. We show that security is achieved through digitally signing and validating all Data packets between NDN gateways and preventing unauthorized 
gateways from injecting arbitrary data into the network. We also demonstrate NDN's ability to prevent masquerading, DoS, and 
replay attacks between different network segments connected through NDN.

\section{Background and Related Work}
    \begin{figure*}[!ht]
        \centering
        \includegraphics[width=0.6\textwidth]{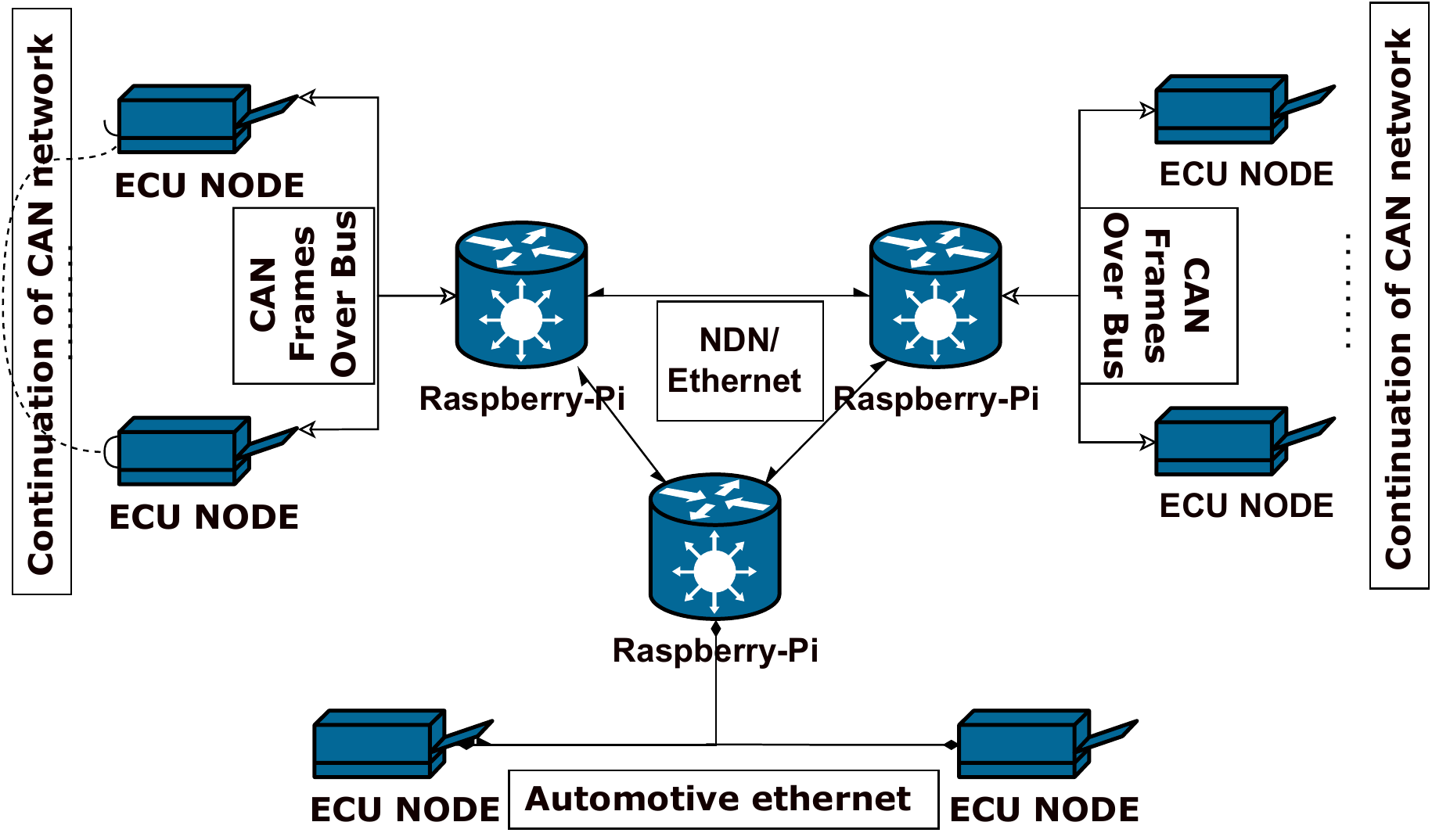}
        \caption{In-vehicle communication architecture with NDN gateways}
        \label{fig:testbed}
    \end{figure*}
\subsection{Controller Area Network}
The Controller Area Network (CAN) is the most common in-vehicle communication network technology used in many types of vehicles such as trucks, heavy equipment, military combat, and support systems; it is also used in sectors other than automobiles, such as in medical devices, industrial equipment, and aviation. CAN was developed by Robert Bosch GmbH in the early 1980s and has been the standard in cars and light trucks since 1996. CAN consists of a broadcast message-based protocol, which supports distributed real-time control and is robust, low-cost, lightweight, simple to configure, and has excellent fault tolerance capabilities. CAN is defined by the ISO 11898-X set of standards, with ISO 11898-2 defining the most common standard, high-speed CAN, which supports data rates up to 1 Mbps with a bus length of 40 m. ISO 11898-3 defines a fault-tolerant CAN, with data rates up to 125 Kbps, which allows transmission of data asymmetrically over one single bus line in case of an electrical failure in another line. Bosch introduced CAN Flexible Data-Rate (CAN-FD) in 2012, which supports data rates faster than 1 Mbps and increases the payload from 8 bytes to 64 bytes \cite{novais2016survey}. CAN is used in automobiles because it is extremely reliable and message loss is extremely rare. It also ensures that a message will be sent and received within a deadline mandated by the automotive industry. CAN messages do not have a source or destination address, instead, nodes interpret whether a message is intended for them based on identifiers called arbitration identifiers (AIDs) that represent what type of data the message holds. In general, there are two formats of CAN: the standard format, which has an 11-bit identifier, and the extended format, which has a 29-bit identifier.

While CAN is lightweight and reliable, it has limited security features, with no encryption or authentication, and has been proven to be exploitable via direct access or remotely. When CAN was designed in the 1980's, the only way to infiltrate or hack a vehicle was to physically break into the vehicle, plug into its on-board diagnostics (OBD) port, and then perform an eavesdropping, injection, or more sophisticated attacks. Unfortunately, modern vehicles exhibit many additional remote attack surfaces, such as keyless entry, Bluetooth, CD, Radio, WiFi, and the Internet. It has been demonstrated that an attacker may infiltrate through one of the aforementioned attack surfaces to reach the CAN controller, which can then be updated with malicious software that would then allow an attacker to send CAN signals to affect functionality such as steering, braking, accelerations, etc.  For example, \cite{miller2014survey, miller2015remote} successfully demonstrated a masquerade attack on a Jeep Cherokee, where a flaw in the vehicle's Internet-connected entertainment system let anyone with the vehicle's IP address to gain access from anywhere in the world and send CAN frames that affect the engine and transmission. In the future, it is expected that the number of attack surfaces will increase even more as the demand for connectivity and vehicle-to-vehicle features increases.

\subsection{Automotive Communication Architectures}

Much like personal computers before them, vehicles are now the subject of innumerable technological advancements both implemented and under consideration for the future. Similar to computers, not a single institution is responsible for all of these innovations. Moreover, there is an active discussion about what the best direction is for future implementations. Because of this, the consensus is that the era of simple, homogeneous vehicles networks is not only coming to an end but will be replaced with faster networks such as automotive Ethernet, which will enable innovation as well as enable sophisticated security mechanisms. For cost and other reasons, vehicles in the near future will contain a mix of networking technologies including CAN and automotive Ethernet. In place of a single or several CAN segments, the future is likely to see CAN isolated to real-time applications such as engine function, while others will be replaced by technologies such as automotive Ethernet. Further, cars have already begun to include external interfaces such as V2V or V2I communicating with short-range WiFi. 
As cars become more connected and incorporate different link-layer technologies, the complexity of the current model makes it impossible to integrate them efficiently due to the cost and size of wiring. Because of this, an alternative model must be found.

Centralized architectures consolidate the number of ECUs into few, powerful compute units \cite{centralNetWorks}
Instead of having several discrete ECUs, they are consolidated into one or more centralized ECUs. The reduced number of ECUs reduces cost, makes software updates easier. It also allows for more rapid experimentation and the introduction of new features. It also reduces the need for complex security due to the low number of ECUs. 

On the other hand, a single ECU can be a single point of failure. In addition, the amount of data that needs to be moved and processed by the central ECU is enormous. Further, since all communication must go through the central ECU, it presents a convenient location for attackers.

Zonal is another type of architecture being explored for in-vehicle communication. In a Zonal Architecture, a network consists of a set of segments interconnected by gateways that are networked together. The segments could be created in any number of ways, but the primary segmentation would be based on link-layer technologies, function, and security considerations. Because segments are partially based on the link layer, the operation of the networking technologies would not have to change inside the segment. Additionally, by creating segments for specific purposes (i.e. engine function, door function, tire function) each segment would be able to function without interference from other segments, minimizing network congestion and security vulnerabilities from external communication. Segmenting implements several foundational security principles that are missing in-vehicle architectures. These include separation of privilege, separation of duty and least privilege. Also, by creating segments, one can customize security for each segment individually. This allows vehicle designers to choose how much security to incorporate into a given segment based solely on that segment's needs instead of being constrained by the security necessities of the vehicle as a whole. Further, the gateways provide the sole location to implement extraneous functionality, including security measures such as intrusion detection on their segment (this is discussed below) allowing each network segment to function unimpaired.

One disadvantage of this architecture is that each segment is vulnerable to attacks from within. However, the presence of the gateway provides an ideal location to implement anomaly detection or a firewall that is specific to the link layer it is monitoring, such as with CAN. Substantial work has been done in this area to secure CAN.
For example, \cite{longari2020cannolo} introduced CANnolo, a reconstruction-based unsupervised IDS that utilizes Long Short-Term Memory (LSTM) autoencoders to identify anomalous CAN messages. Other popular CAN IDS approaches include rule-based, clock-based, and signal-basd approaches \cite{tariq2020can, halder2020coids, verma2021can}. Additionally, \cite{humayed2020cansentry} proposed CANSentry, a firewall that can be deployed between potentially malicious CAN nodes and the bus to protect against denial of service attacks.

Each gateway must be computationally capable of performing a set of tasks ranging in complexity, including anomaly detection and the ability to filter requests and responses from various segments. In this design, the gateway acts as a firewall and a translator between segments and other gateways. 


\section{NDN for Automotive Communication}

\subsection{Named Data Networking}

Named Data Networking (NDN)~\cite{Jacobson:2009:NNC:1658939.1658941} is 
a future Internet architecture that is designed to fetch named and signed Data packets instead of delivering IP packets to destination hosts.
NDN allows applications to use semantically meaningful names for retrieving data. For example, when an Anti-lock Breaking System (ABS) needs the current rotation speed of the left rear wheel, it can express an Interest by the name ``/vehicle/chassis/RearAxle/LeftWheel/rotationSpeed/rpm". The network forwards the request towards a potential Data producer, which is the sensor in the wheel in the case. Once the request reaches the sensor, it digitally signs and returns the Data to the requester.


From this example, we can have an idea about the basic 
architecture of NDN: A requester sends Interest packets to fetch Data packets. An Interest packet contains a semantic meaningful name that helps fetch a Data packet by exact or partial name matching. Moreover, a Data packet can be fetched directly from the data producer or an in-network cache.

While NDN provides many important features like caching, Data reuse, etc., one of the primary advantages of using NDN is security. An NDN Data packet is always digitally signed irrespective of its retrieving location (e.g., directly from producer or cache). This signature binds the name to the content.
A consumer can verify the signature and therefore, trust the Data packet. 
Besides, NDN Data packets are immutable and so changing the content will create a new data packet with a new name and thus it can be distinguished from the original one.

NDN names also allow us to introduce trust schema~\cite{yu2015schematizing} that defines if Data packets have the necessary signature to be considered ``trusted".
In NDN, a Data packet contains a key name (in KeyLocator field of a data packet) to point out which public key a consumer should fetch to verify the signature of that particular data packet.
A trust schema specifies whether that key was a legitimate key to sign that particular Data packet or not.
For example, we can specify that ``/vehicle/chassis/RearAxle/LeftWheel/rotationSpeed/rpm" needs to be signed by ``/vehicle/chassis/RearAxle/LeftWheel/KEY"(i.e.,We can generalize this by prefix-matching). Therefore if any malicious 
ECU tries to reply for that particular Interest and signs with a different key (``/vehicle/maliciousECU/KEY"), the consumer can easily identify that and ignore the data packet.
A trust schema ensures that if a particular component somehow gets compromised, it still cannot affect the overall integrity of communication for the rest of the components.

NDN also supports data confidentiality and access control using encryption and Name-based Access Control (NAC) scheme~\cite{zhang2018nac}.
Using NAC, a producer encrypts content at the time of production and distributes the decryption key automatically only to the desired consumer(s)
Thus NDN can prevent unauthorized access to any private data.

Therefore, by adapting NDN’s security mechanism, Data packets ensure content integrity, authenticity, and (if encrypted) confidentiality irrespective of how the Data packet is retrieved.

In a vehicle network with NDN gateways, each gateway could be configured with information about the types of requests that ECUs within its segment could make. These names can be used for making requests. Gateways can control which names they forward to the other segments based on the request names.
Note that information relating to what data might be requested from a segment and the names that an ECU would forward
could easily be configured into the gateways before deployment and updated later either dynamically or by software updates from the manufacturer. Table \ref{tab:signing_overhead} summarizes how NDN addresses most common vulnerabilities present in today's automotive networks. 

\begin{table}
\begin{center}
\caption{Vehicle Attack And Mitigation using NDN}

\begin{tabular}{|c|c|} 
 \hline
CAN Vulnerability & NDN Mitigation \\ [0.5ex] 
 \hline\hline
 Masquerading& Signing \\ 
 \hline
 Eavesdropping & None \\
 \hline
 Replay Attack  & No unsolicited data \\
 \hline
 Injection Attack & Signing \& Protocol Design  \\
 \hline
 Denial of service  & No unsolicited data\\ [1ex] 
 \hline
 Bus-off attack & Not applicable \\
 \hline
\end{tabular}
 \label{tab:signing_overhead}

\end{center}
\end{table}

\section {Threat Model}

In this section we describe the threat model we use in this paper. We consider four attacks; for each attack we describe the attack, the conditions that enable it, and the potential harm from the attack.

\textbf {Unsolicited Traffic.}
In this attack, a malicious ECU sends unsolicited traffic to fake sensor readings, or trigger an unwanted behavior by other ECUs, or affect the vehicle in some other negative way. Such an attack could be mounted by a compromised ECU, or a malicious counterfeit component installed in the vehicle. The potential harm to the vehicle depends on the messages sent and varies from displaying fake data to triggering operation that causes physical harm to the vehicle, its occupants, or property.

\textbf {Masquerading.}
In this attack, a malicious ECU sends data pretending to be a legitimate ECU. The conditions enabling such an attack are similar to unsolicited traffic, but with a higher probability of harm if there are no checks or if all checks fail at the receiving ECU and it accepts the masqueraded messages.

\textbf {Replay Attacks.}
In this attack, a malicious ECU captures legitimate traffic and replays it at a later time. Any compromised or counterfeit ECU can potentially mount a replay attack. The harm done depends on the replayed message (similar to unsolicited and masquerading attacks) and the safeguards the receiving ECU has in place to recognize duplicate messages. The harm also depends on whether the attack is mounted on a physical actuator or the logical operation of the vehicle, and can range from making the vehicle inoperable to causing harm to the occupants or property.

\textbf {Denial of Service (DoS) Attacks.}
A DoS attack happens when a malicious ECU floods the network with unwanted traffic denying service to other ECUs. An ECU can attack the network, in which case all ECUs are under DoS, a specific ECU, or both. This can happen when an ECU is compromised, or a counterfeit ECU is installed in the vehicle. Potential harm depends on when the attack is triggered and can range from a stalled vehicle to physical harm if communication is denied at a critical moment (e.g., right before the application of the brakes). 

Our architecture addresses all these attacks as we describe in the Evaluation section. Worst case scenario an attack will be limited to the CAN segment a malicious ECU is on, but will be quickly detected. If the malicious ECU is on the NDN part of the network the attack is either not possible or will be quickly neutralized.

\label{evaluation}

\section{Experimental Setup}

In this section, we begin by presenting our testbed implementation followed by our evaluation results. We separate the results into those demonstrating security and those demonstrating performance. The security results were performed on a laptop since they only demonstrate the security properties of NDN. Performance results were performed on the actual testbed. However, note that the performance results are for illustration purposes only, since the current NDN implementation is not optimized for performance.

\subsection{Testbed Description}
\begin{figure}[!ht]
        \centering
        \includegraphics[trim={0 4cm 0 0},width=0.45\textwidth]{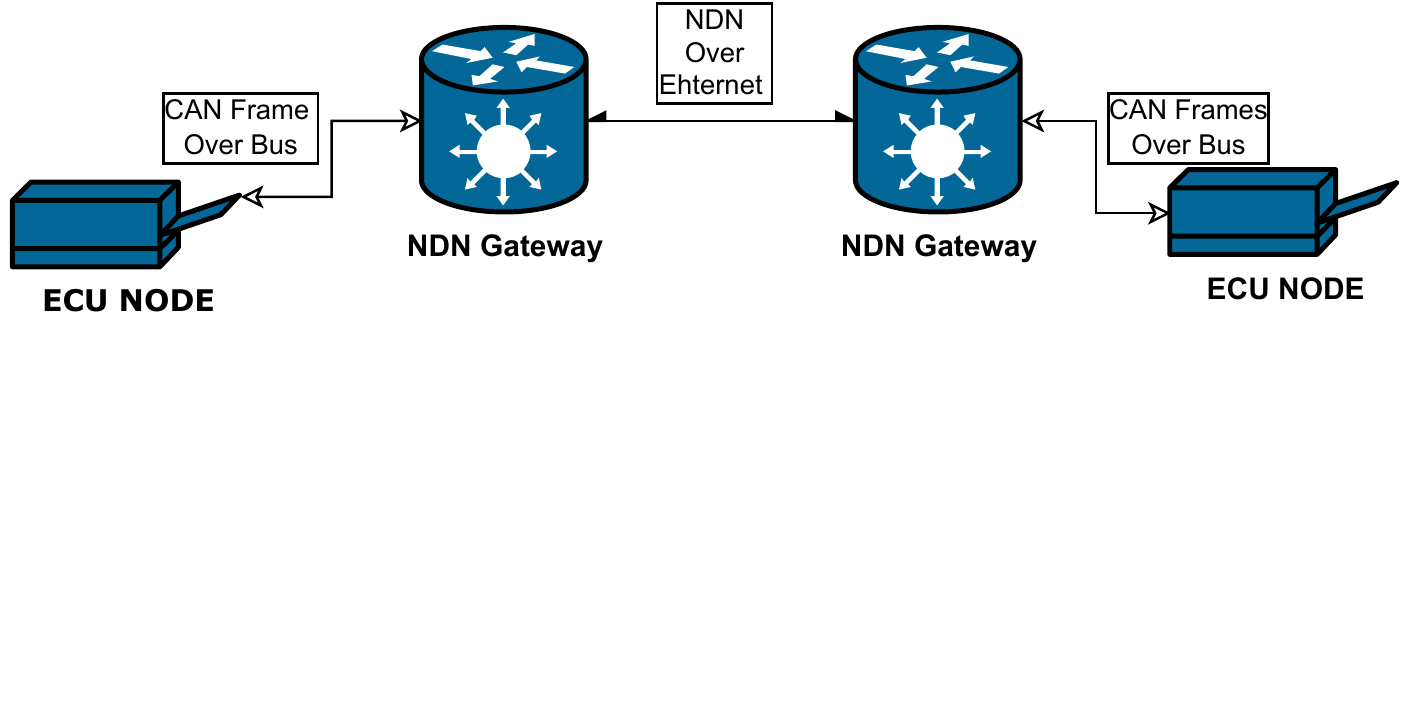}
        \caption{Testbed Implementation}
        \label{fig:testbed}
\end{figure}

We developed a simple benchtop testbed as a proof-of-concept to test our claims. Figure \ref{fig:testbed} shows the testbed. The testbed emulates a generic architecture with two high-speed nodes communicating over NDN which we refer to as gateways, connecting two CAN segments, which represent legacy networks.  This testbed demonstrates the high-speed communication between different CAN segments by using NDN, further it demonstrates how content is secured by passing it through NDN. Communication over the individual CAN segments is not secured. Such scenarios appear in both zonal and centralized architectures and in general any architecture where there is a mix of legacy and new, high-speed networks.   In our implementation, CAN segments are connected to NDN gateways that are connected over standard Ethernet, which we use as a stand-in for Automotive Ethernet. The testbed consists of four Raspberry Pi 3s running Ubuntu Server 20.04.3 LTS. Attached to each Raspberry PI is a Pi CAN HAT\cite{CANHAT} capable of transmitting and receiving CAN messages. 

The testbed has two PIs acting as NDN gateways that are capable of communicating using both Ethernet and CAN. These nodes are used to tunnel CAN messages over NDN, and support the Interest/Data functionality. In our design, we only transmit CAN frames between the two segments, but there is nothing to prevent us from integrating communication over other link layer technologies such as automotive Ethernet.
The other two Pis represent CAN ECUs.
These are connected to the NDN gateways over CAN using CAN HATs. These Pis are used to replay CAN data from a trace\cite{Verma2020ROADTR},
so they emulate the ECUs that produced the original trace.   

Public keys were exchanged before the tests for the purpose of signing and validation between the NDN gateways. 
While both gateways perform the same functions, we designate the left gateway as the ``Producer" and the right gateway as the ``Consumer". The consumer requests content by sending Interests and the producer responds to the requests using Data packets.

\section{Evaluation}

This section describes both our security and performance evaluation experiments. The experiments were done in a mix of testbeds, a laptop and our automotive testbed with four Raspberry Pis. We used a laptop for convenience when it would not affect the experiment conclusions. For each experiment we state explicitly where it was done. 

\subsection{NDN Security Evaluation}




Recall that NDN provides security features not found in CAN.  NDN producers sign the Data packets they send using their private key and when a Data packet is received, NDN consumers validate them in a three-step process: (a) verify that the signatures are of the same type (e.g., SHA256 with RSA, ECDSA, or HMAC), (b) ensure that the name of the signer matches the expected name of the key, and (c) if the first two checks pass, check the signature for validity. An attacker may forge a Data packet to pass the first two checks but will fail the final step since the signature is not valid.


\subsubsection{Unsolicited Data}
In this attack an attacker attempts to send a Data packet to a consumer without the consumer sending an Interest. Unsolicited Data attacks are impossible to execute in NDN. An attacker can generate a Data packet, or replay an old Data packet, but without an Interest there is no path in the network for the Data packet to follow. NDN will only forward a Data packet if preceded by an Interest that created state in the network. We did not do an experiment at the NDN level to address this attack. However, we note that there are attacks possible at the link layer or IP layer that may deliver a Data packet to an NDN consumer, but these would fail due to lack of forwarding state or an invalid signature. These attacks are covered in the next experiment.


\subsubsection{Impersonation/Masquerading}
In this attack, the attacker sends Data packets with an invalid signature in response to Interest packets. Such an attack may be carried out if an attacker is monitoring Interest traffic and injecting Data packets, or if it impersonates a valid producer and attracts Interests. In this case Data packets produced by the attacker will reach the consumer. However, as we noted earlier, the Data packets will be rejected due to signature failure. We carried out this experiment on the testbed. We assigned signing keys of the same type to the NDN gateways so they can sign and validate the Data packets they send and receive. Then we executed an attack where the attacker signed its Data packets with its own key but changed the name of the key to the one expected by the signature validator. In all cases, the Data packet failed to be validated because the signatures did not match. We therefore conclude that an attacker who does not have the correct private key may push through Data packets if timed properly and in response to Interests, but the receiver will always reject them.

\subsubsection{Replay attack}
For the next attack we consider the ability of an attacker to capture messages on the bus and replay them to the consumer. The NDN architecture guards against such a scenario natively, by making the current timestamp as part of the request (or a monotonically increasing sequence number). This has a similar effect to a nonce that is unique for every message request and receipt. This nonce is then signed as part of the response to the data. A replayed message with an older nonce is easy to detect and drop,  even if it passes validation of the signature.

\subsubsection{DoS using Interest Packets}

Finally we consider the case of a DoS attack by an attacker flooding the gateway with Interests. Performing a DoS attack using Interest packets is much easier than DoS attacks with Data packets since unlike Data packets Interest packets can be unsolicited. An attacker can simply flood a Producer with Interests. There are two types of attacks using Interest packets: (a) an attack with Interests asking for the same content, and (b) and an attack with Interests asking for different, unique content each time. The first type of attack will not reach the Producer since the first NDN forwarder consolidates all Interests asking for the same content. The second type of attack, however, will result in all Interests reaching the Producer. Most modern NDN implementations can guard against this case by monitoring the rate of incoming Interests and/or congestion, and issuing NACKs (therefore denying the satisfaction of the Interest) to clients, before their links are overwhelmed.

We tested both types of Interest attacks in our testbed. For the attack requesting the same content, attacking Interests were consolidated as expected and did not overwhelm the Producer. Only the first Interest reached the producer and retrieved the content, which was subsequently cached at the forwarders, so future Interests retrieved content from the nearest cache. Thus, this attack was effectively mitigated because it depends on when the cached content expired at the forwarder. Data packets are cached based on the Freshness attribute of the content, and if long enough this will eliminate the attack.

In the second Interest attack an attacker makes repeated requests for new data, which reach the Producer. This is a DoS attack on the producer, who may get overwhelmed by a large number of Interests. Note that the Interests can request published or non-published data, but the effect on the Producer is similar and depends on the Producer implementation. Bogus Interests requesting existing data can also overwhelm the return path. NDN implementations have the ability to detect such attacks by keeping track of the rate of unsatisfied Interests. Thus, all a producer has to do is not respond to Interests considered as an attack, and NDN will notice the high rate of unsatisfied Interests. There is currently no mitigation mechanism in NDN, but this is a topic of future work. We tried these attacks in our testbed, but they did not succeed, because the testbed was not capable of generating Interests at a high enough rate. The Producer was able to respond to all Interests, and we never saw the Producer's CPU usage exceed 20\%. However, this is an artifact of our testbed, and such DoS attacks are feasible in other networks, especially if they contain more powerful or enough attackers. We will investigate mitigation techniques in future work.


\subsection{Performance Evaluation}

In this section, we evaluate the latency overhead incurred by NDN. We again emphasize that these measurements are for illustration purposes only, since the current NFD implementation was not optimized for performance. We measure the security delays separately to gain better insight into the overhead of NDN. We measured latency at each leg in the testbed separately to account for asynchronous transmission and make sure that our calculations did not contain any of the time between when content was generated and signed, and when it was requested.
All of our tests follow the same format. CAN messages are read from a CAN log containing real CAN traffic, and sent over the bus by an ECU. An NDN gateway, the producer, reads these messages from the CAN bus, and then creates and signs each Data packet with its private key before publishing them. A second NDN gateway, the consumer, sends an NDN Interest for the producer Data. The consumer receives and validates received Data packets before converting them back into CAN messages and passing them on its bus to a final ECU that receives them. 

Our tests were designed to measure the overhead of NDN operations. We do not show the end-to-end time for any of the CAN bus transmissions since CAN is not the object of our investigation. We ran each experiment 1000 times and took the average of the delays in seconds at the producing and consuming gateways separately. Then we ran each experiment again, and the independent averages of the two runs were averaged. We preformed two experiments, one where NDN Data packets were generated by request and another where the Data packets were generated when the CAN messages were received and buffered until a request for that data was made.

\begin{figure}
    \centering
    \includegraphics[width=0.49\textwidth]{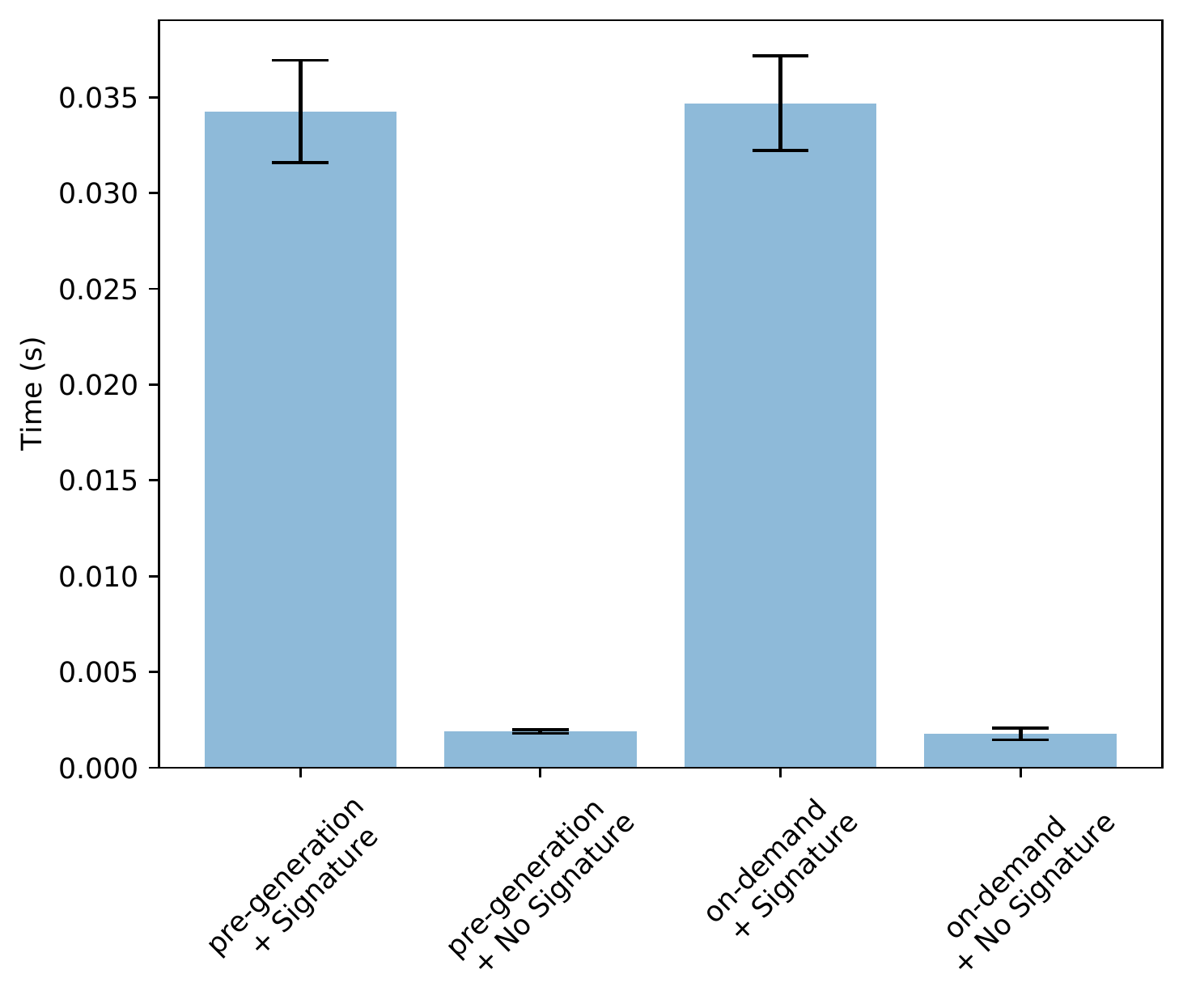}
    \caption{Producer-side delays}
    \label{fig:producer}
\end{figure}

\begin{figure}
    \centering
    \includegraphics[width=0.49\textwidth]{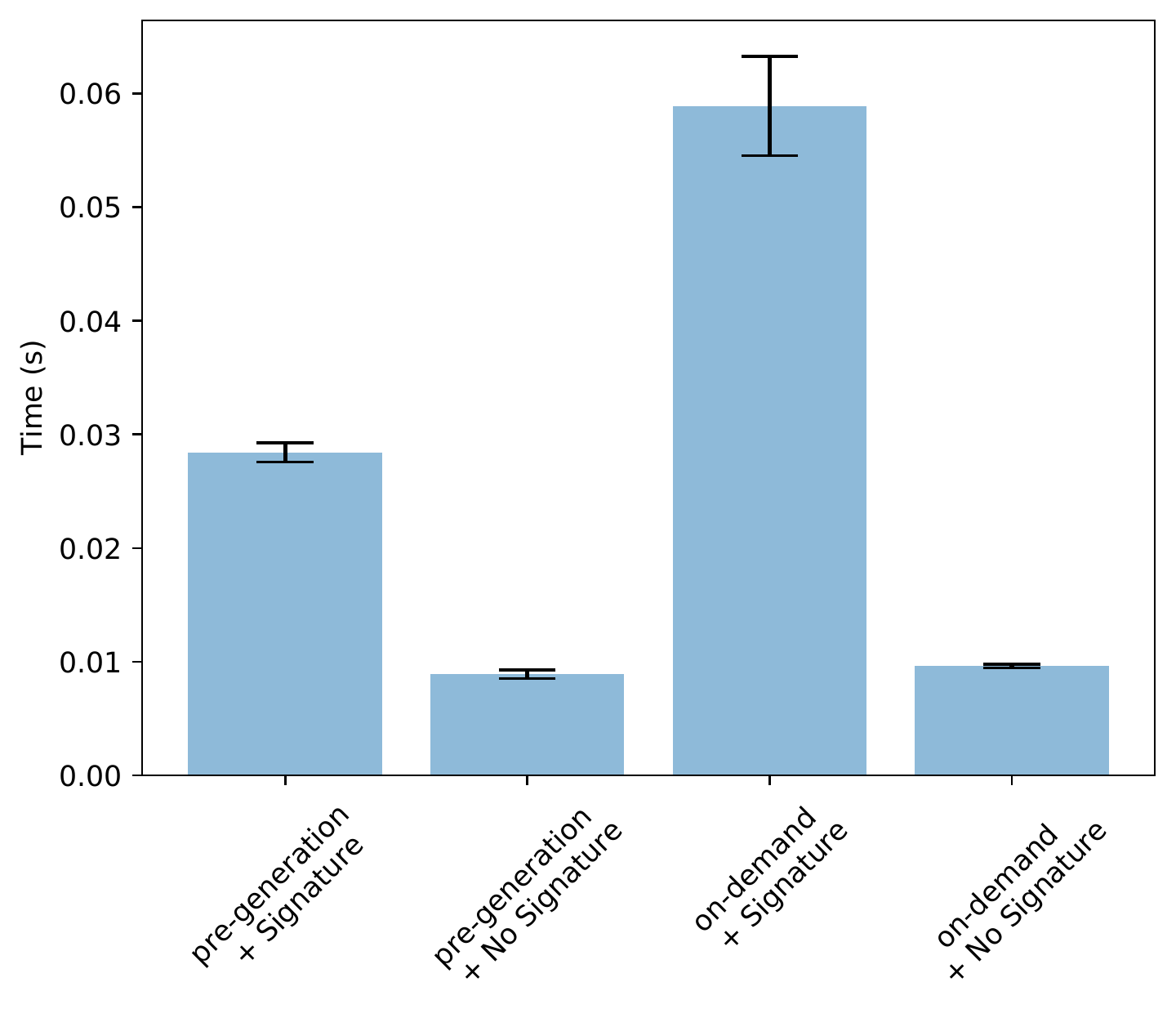}
    \caption{Consumer-side delays}
    \label{fig:consumer}
\end{figure}


\subsubsection{Interest\//Data Exchange with On-Demand Generated NDN Packets}

In this experiment NDN Data packets were generated on-demand after receiving an Interest packet. On-demand generation includes both creating and signing a Data packet. The values for the consumer in Table \ref{tab:on-demand data} represent the total latency measured at the consumer NDN gateway. This includes the time (a) for the consumer to generate and send an Interest, (b) the Producer to receive the Interest, (c) create and sign the requested Data packet, (d) transmit it back to the consumer, (e) the time for the consumer to receive it (f) and the time it takes the consumer to validate its signature. For this test the results for the producer gateway are the sums of parts (c) (d). Note that we ran this test with and without signatures and validation to isolate the security overhead. 


   \begin{table}
   \begin{center}
    \caption{Time for packet creation and validation for pre-generated data (Seconds)}

\begin{tabular}{|p{1cm}|p{2cm}|p{2cm}|} 
 \hline
 Overhead & Producer & Consumer\\ [0.5ex] 
 \hline\hline
 With Signing & Mean: 0.0342, STDEV: 0.0027 & Mean: 0.0284, STDEV: .0008 \\
 \hline
 Without Signing & Mean: 0.0019, STDEV: .0001 & Mean: 0.0089, STDEV: .0004 \\
 \hline
\end{tabular}
 \label{tab:pre-generated data}
\end{center}  
   \end{table}
   
   \begin{table}
   \begin{center}

\caption{Time for packet creation and validation for on-demand data (Seconds)}
\begin{tabular}{|p{2cm}|p{2cm}|p{2cm}|} 
 \hline
 Overhead & Producer & Consumer\\ [0.5ex] 
 \hline\hline
 With Signing and Validation &  Mean: 0.0312, STDEV:0.0025 & Mean: 0.0589, STDEV:0.0044 \\
 \hline
 Without Signing and Validation & Mean: .0013, STDEV:0.0001 & Mean: .0096, STDEV:0.0002 \\
 \hline
\end{tabular}
\label{tab:on-demand data}
\end{center}  
   \end{table}


   


\subsubsection{Interest\//Data Exchange with Pre-Generated NDN Packets}

We performed a second experiment where data packets were generated ahead of time before an Interest arrived. The goal was to measure the savings of pre-generating Data packets in anticipation of an Interest. Such Interests are often easy to anticipate if, for example, we know that an ECU is interested in receiving continuous updates. For this experiment the Producer creates a Data packet as soon as it receives a CAN message. The Data packet is then buffered and ready to transmit upon receiving an Interest. 
The values for the consumer in Table \ref{tab:pre-generated data} represent the time (a) for the consumer NDN gateway to generate the Interest, (b) transmit it to the producer,(c) the producer to publish the respective Data packet (d) and the consumer to receive and validate it. The values for the producer are no longer included in the consumer overhead calculations because its work is done at the time CAN frames are received. The values for the producer represent the time it takes to make an NDN packet and place it in a buffer to later be published when an associated interest is made. We found that by pre-generating and signing Data packets we were able to reduce response time by a factor of 2.

\section{Conclusions}

In this paper we presented NDN, a communication architecture that provides security by design by signing all content transmitted in the network. We demonstrated the security features of NDN through a number of attacks. NDN is well suited for high-speed, secure communication in in-vehicle networks, interconnecting high-speed and legacy segments. NDN does not secure legacy networks such as CAN, but ensures that communication between such segments that passes through NDN is secure.

In the future, we intend to expand our work in several ways. We will implement richer topologies that include segments with other link-layer technologies, such as automotive Ethernet, LIN, and Zigbee. We will include more realistic scenarios with mixed traffic from different networks. We will implement and test techniques to mitigate various attacks. Finally, we will adopt the Vehicle Signal Specification (VSS)~\cite{VSS} naming scheme for NDN.



\bibliographystyle{IEEEtran}
\bibliography{refs.bib}

\begin{thebibliography}{10}
\providecommand{\url}[1]{#1}
\csname url@samestyle\endcsname
\providecommand{\newblock}{\relax}
\providecommand{\bibinfo}[2]{#2}
\providecommand{\BIBentrySTDinterwordspacing}{\spaceskip=0pt\relax}
\providecommand{\BIBentryALTinterwordstretchfactor}{4}
\providecommand{\BIBentryALTinterwordspacing}{\spaceskip=\fontdimen2\font plus
\BIBentryALTinterwordstretchfactor\fontdimen3\font minus
  \fontdimen4\font\relax}
\providecommand{\BIBforeignlanguage}[2]{{%
\expandafter\ifx\csname l@#1\endcsname\relax
\typeout{** WARNING: IEEEtran.bst: No hyphenation pattern has been}%
\typeout{** loaded for the language `#1'. Using the pattern for}%
\typeout{** the default language instead.}%
\else
\language=\csname l@#1\endcsname
\fi
#2}}
\providecommand{\BIBdecl}{\relax}
\BIBdecl

\bibitem{ullah2021}
M.~A. Ullah, S.~Ghafoor, M.~Rogers, and S.~Prowell, ``Seccan: A practical
  secure control area network for automobiles,'' 02 2021.

\bibitem{9644625}
C.~Papadopoulos, A.~Afanasyev, and S.~Shannigrahi, ``A name-based secure
  communications architecture for vehicular networks,'' in \emph{2021 IEEE
  Vehicular Networking Conference (VNC)}, 2021, pp. 178--181.

\bibitem{novais2016survey}
P.~Novais and S.~Konomi, ``A survey on vehicular communication technologies,''
  in \emph{Proc. 12th Int. Conf. Intell. Environ.}, 2016, p. 308.

\bibitem{miller2014survey}
C.~Miller and C.~Valasek, ``A survey of remote automotive attack surfaces,''
  \emph{black hat USA}, vol. 2014, p.~94, 2014.

\bibitem{miller2015remote}
------, ``Remote exploitation of an unaltered passenger vehicle,'' \emph{Black
  Hat USA}, vol. 2015, no. S 91, 2015.

\bibitem{centralNetWorks}
\BIBentryALTinterwordspacing
J.~Scobie, ``{The Future of Mobility: A Centralized Vehicle Architecture},''
  \emph{Electronic Design}, Jun 2021. [Online]. Available:
  \url{https://www.electronicdesign.com/markets/automotive/article/21149210/arm-the-future-of-mobility-a-centralized-vehicle-architecture}
\BIBentrySTDinterwordspacing

\bibitem{longari2020cannolo}
S.~Longari, D.~H.~N. Valcarcel, M.~Zago, M.~Carminati, and S.~Zanero,
  ``Cannolo: An anomaly detection system based on lstm autoencoders for
  controller area network,'' \emph{IEEE Transactions on Network and Service
  Management}, vol.~18, no.~2, pp. 1913--1924, 2020.

\bibitem{tariq2020can}
S.~Tariq, S.~Lee, H.~K. Kim, and S.~S. Woo, ``Can-adf: The controller area
  network attack detection framework,'' \emph{Computers \& Security}, vol.~94,
  p. 101857, 2020.

\bibitem{halder2020coids}
S.~Halder, M.~Conti, and S.~K. Das, ``Coids: A clock offset based intrusion
  detection system for controller area networks,'' in \emph{Proceedings of the
  21st International Conference on Distributed Computing and Networking}, 2020,
  pp. 1--10.

\bibitem{verma2021can}
M.~E. Verma, R.~A. Bridges, J.~J. Sosnowski, S.~C. Hollifield, and M.~D.
  Iannacone, ``Can-d: A modular four-step pipeline for comprehensively decoding
  controller area network data,'' \emph{IEEE Transactions on Vehicular
  Technology}, 2021.

\bibitem{humayed2020cansentry}
A.~Humayed, F.~Li, J.~Lin, and B.~Luo, ``Cansentry: Securing can-based
  cyber-physical systems against denial and spoofing attacks,'' in
  \emph{European Symposium on Research in Computer Security}.\hskip 1em plus
  0.5em minus 0.4em\relax Springer, 2020, pp. 153--173.

\bibitem{Jacobson:2009:NNC:1658939.1658941}
\BIBentryALTinterwordspacing
V.~Jacobson, D.~K. Smetters, J.~D. Thornton, M.~F. Plass, N.~H. Briggs, and
  R.~L. Braynard, ``Networking named content,'' in \emph{Proceedings of the 5th
  International Conference on Emerging Networking Experiments and
  Technologies}, ser. CoNEXT '09.\hskip 1em plus 0.5em minus 0.4em\relax New
  York, NY, USA: ACM, 2009, pp. 1--12. [Online]. Available:
  \url{http://doi.acm.org/10.1145/1658939.1658941}
\BIBentrySTDinterwordspacing

\bibitem{yu2015schematizing}
Y.~Yu, A.~Afanasyev, D.~Clark, V.~Jacobson, L.~Zhang \emph{et~al.},
  ``Schematizing and automating trust in named data networking,'' in
  \emph{Proc. 2nd ACM ICN Conf.}, 2015, pp. 1--10.

\bibitem{zhang2018nac}
Z.~Zhang, Y.~Yu, S.~K. Ramani, A.~Afanasyev, and L.~Zhang, ``Nac: Automating
  access control via named data,'' in \emph{MILCOM 2018-2018 IEEE Military
  Communications Conference (MILCOM)}.\hskip 1em plus 0.5em minus 0.4em\relax
  IEEE, 2018, pp. 626--633.

\bibitem{CANHAT}
\BIBentryALTinterwordspacing
``{PiCAN2 - Controller Area Network (CAN) Interface for Raspberry Pi},'' Mar
  2022, [Online; accessed 31. Mar. 2022]. [Online]. Available:
  \url{https://copperhilltech.com/pican2-controller-area-network-can-interface-for-raspberry-pi}
\BIBentrySTDinterwordspacing

\bibitem{Verma2020ROADTR}
M.~E. Verma, M.~D. Iannacone, R.~A. Bridges, S.~C. Hollifield, B.~Kay, and
  F.~L. Combs, ``Road: The real ornl automotive dynamometer controller area
  network intrusion detection dataset (with a comprehensive can ids dataset
  survey \& guide),'' \emph{ArXiv}, vol. abs/2012.14600, 2020.

\bibitem{VSS}
\BIBentryALTinterwordspacing
``{Vehicle Signal Specification (VSS)/Vehicle Data Spec - Auto},'' Mar 2022,
  [Online; accessed 31. Mar. 2022]. [Online]. Available:
  \url{https://www.w3.org/auto/wg/wiki/Vehicle\_Signal\_Specification\_(VSS)/Vehicle\_Data\_Spec}
\BIBentrySTDinterwordspacing

\end{thebibliography}

\end{document}